# Transport dynamics study of laser-accelerated proton beams and design of double achromatic beam translation system

J. G. Zhu[1], *H. Y. Lu[1], Y. Zhao[1], M. F. Lai[1], Y. L. Gu[1], #S. X. Xu[2] and †C. T. Zhou[1]

[1] *Shenzhen Key Laboratory of Ultraintense Laser and Advanced Material Technology, Center for Advanced Material Diagnostic Technology, and College of Engineering Physics, Shenzhen Technology University, Shenzhen 518118, China*

[2] *Shenzhen Key Lab of Micro–Nano Photonic Information Technology, Key Laboratory of Optoelectronic Devices and Systems of Ministry of Education and Guangdong Province, College of Physics and Optoelectronic Engineering, Shenzhen University, Shenzhen 518060, China*

Correspondence to: *luhaiyang@sztu.edu.cn，

†zcangtao@sztu.edu.cn,

#shxxu@szu.edu.cn;

Proton beams generated from laser acceleration show merit for their unique spatial (micron-size) and temporal (picosecond) properties, which make them desirable for many potential applications. However, the large energy spread and divergence angle make it difficult to maintain these beam properties after delivery. This hinders the wide application of laser acceleration. In this paper, we design a double achromatic beam translation system (DABTS), based on weak-focusing magnets, to realize achromatic transmission in both the horizontal and vertical directions, and compress the bunch length of the delivered proton beam at the same time. We make use of fringe angles and special steering magnets to effectively reduce the influence of chromatic aberrations and high-order nonlinear terms and realize close to ideal point-to-point optics. We believe this work contributes to the ongoing effort to apply laser accelerators to a variety of fields.

## I. INTRODUCTION

The interaction of high-intensity ultra-short lasers with materials generates plentiful transient phenomena, which may open new horizons in scientific research, such as particle acceleration. Large electric field gradients (>100 GV/m) can be created in a laser accelerator, significantly exceeding the accelerating fields (a few tens of MeV/m) in conventional accelerators which are limited to material breakdown. These high gradients hold the promise of developing more compact and economical accelerators. Laser accelerators have also attracted much attention for their unique properties, including short temporal duration (several picosecond) [1], small beam spot (several microns), high brightness, small emittance, large divergence angle and broad energy spread [2,3]. As effectively point sources, they enable potential applications for precise time-resolved imaging of fast transient phenomena [4], emulating space conditions [5], cancer therapy [6] and fast ignition of fusion cores [7].

The most widely employed laser proton acceleration mechanism is target normal sheath acceleration (TNSA). After laser irradiation, a strong electrostatic sheath field is created at the rear surface of the target by hot electrons, and protons can be accelerated by this field to tens of MeV, but with nearly 100% energy spread [8–10]. Radiation pressure acceleration (RPA) [11–15] and break-out afterburner (BOA) [16] are other important mechanisms. Proton energies up to nearly 100 MeV have been achieved in the RPA and BOA regimes [17,18].

In addition to energy, different practical applications have diverse requirements on proton beams. For instance, fast ignition of fusion cores needs beams with high brightness (hundred-micron sizes) and hundred-picosecond bunch length [19], and cancer therapy needs collimated beams. Most practical applications have rigorous requirements for the properties of the ion beams used.

Many scientific questions have been posed regarding the structure of planets, such as the extreme conditions inside Neptune that may cause carbon to form into diamond layers [20]. Study of warm dense matter (WDM) in the laboratory is a key way to investigate these questions. Isochorically-heated WDM can be created using laser-driven proton beams [21,22]. Thanks to the well-known energy deposition curve of protons, a long energy absorption length for samples (tens to hundreds of μm thick) can be controlled precisely with the proton energy spectrum [23]. The continuous energy spectrum of laser-accelerated proton beams can indeed be an advantage in such applications. However, there are three obstacles. Firstly, to reach the isochoric conditions and obtain high temperatures, the secondary heated targets are usually located no further than 1 mm away from the proton source, so that the beam can be reserved on a hundred-micron spatial scale and on a hundred-picosecond time scale [24]. However, hot plasma blow-off and strong electromagnetic pulses (EMP) inevitably interfere with the measurement and diagnostic systems. Secondly, due to the instability of the laser–plasma interaction, laser-accelerated beams usually have significant fluctuations in the energy spectrum, charge and divergence angle [25]. Thirdly, production of a spread-out Bragg peak (SOBP) is key to achieving uniform volumetric heating within a certain depth range with a continuous energy spectrum of the proton beam. The typical spectrum of laser-accelerated protons is exponentially decaying, while the production of SOBP requires an ascending spectrum [26]. Hence accurate energy analysis is needed and the energy spectrum needs to be shaped.

In summary, beam manipulation with a beamline is a prerequisite for precise and controllable generation of a WDM state, and the beamline should be capable of delivering particles possessing certain spatial and temporal properties (on a hundred-micron scale, hundred-picosecond time scale and ascending spectrum). These proton beams are appropriate not only for heating samples to extreme states, but also for time-resolved imaging of fast transient phenomena, and fast ignition.

Research into beam manipulation with a beamline has been carried out widely. As the initial spatial and temporal qualities of laser-driven beams are already correct for the application, the beam transport system must be able to retain these qualities (beam size, emittance, ultrafast beam), which is not easy to achieve. No beamline capable of delivering beams to the application platform with spatial properties as good as at the target has been reported to date.

The most widely used focusing component is the quadrupole lens [27]. Focusing with a magnetic field is energy dependent. No real particle beam is ever strictly monochromatic, therefore the chromatic aberration will inevitably exert an influence on focusing. Focusing components, including quadrupole lenses and solenoid magnets [28,29], have been demonstrated in theory and experiments to be object-to-image point transport systems at the central energy. In practical beam transport, the proton beam source (the laser irradiation point in a laser-driven accelerator) is the object point for each energy. However, focusing differences due to chromatic aberrations cause particles to be imaged at different focal points (namely different longitudinal positions of the image points), leading to a severe degradation of the attainable luminosity. Beam size increases with energy spread [30]. This means that the total beam size cannot be on a hundred-micron scale.

Proton beams generated in laser acceleration have large divergence angle and energy spread. Accurate energy selection requires that protons with the same energy and different angular divergence converge to the same image point at the $x$ (horizontal) axis, while protons with different energies are separated in the $x$ direction. This means the bending magnet must focus particles while deflecting particles in the *x* direction.

A chicane of dipoles is often adopted as an energy selection system [31,32]. As rectangular magnets do not include focusing, only a collimated beam can be analyzed accurately, and the precision of energy selection is determined by the beam size at the entrance. Sector magnets with a uniform magnetic field can focus the beam in the *x* direction and analyze energy at the same time. The focusing is absent in the *y* (vertical) direction, hence specialized focusing elements such as quadrupole lenses are usually added between the beam source and sector magnet to collect and focus. However, particles with different energies will have different longitudinal positions for the image point after being focused by the quadrupole lens, because of chromatic aberrations, and the differences of image point positions will be relayed to the transport in the sector magnet, leading to the requirement for matching-image-point two-dimensional energy analysis to guarantee energy analyzing precision [26]. In addition, energy spectrum shaping should be performed at the image points after the sector magnet in the *y* direction, while chromatic aberrations generate a transverse bow tie profile in the *xy* plane [30], which increases the complexity of spectrum shaping.

Achromatic design, such as a combination of quadrupoles and bending magnets, is often used to eliminate the influence of chromatic aberrations [33], although achromatism can only be achieved in one dimension (horizontal or vertical direction). Chromatic aberrations still impact beam qualities in an achromatic beamline even in the case of an energy spread of 1% and a 1 mrad divergence [34]. The transformation matrix for bending magnets is the same for different energies in linear beam dynamics, while the transformation matrix for quadrupoles is different for different energies. Achromatic design of a beamline is determined at the central energy, but for other energies, the transport cannot be strictly achromatic.

Sector magnets with a constant magnetic field gradient were utilized in circular accelerators like betatrons and the first generation of synchrotrons. The relationship between magnetic field $B$ and radius $r$ can be described as $B = Cr^{-n}$ ($C$ is a constant). When the field index $n$ is between 0 and 1, the magnet can generate focusing forces in the $x$ and $y$ directions simultaneously, which is called weak focusing, in contrast to the case of strong focusing based on quadrupole magnets. Weak focusing was gradually replaced by strong focusing due to the difficulties in economics and construction when focusing proton beams with energy of the order of GeV. However, weak focusing is effective and indeed advantageous in transmitting proton beams at tens to hundreds of MeV. Significantly, a weak-focusing system can realize achromatic transmission in both the $x$ and $y$ directions simultaneously under certain conditions.

A weak-focusing system can focus beams in the $x$ and $y$ directions, and analyze energy in the $x$ direction at the same time. Due to the focusing in the $x$ direction, beams with large divergence angle and energy spread can still be analyzed accurately. Furthermore, chromatic aberrations have little influence in the $y$ direction, hence profiles are almost identical for different energies, which is convenient for spectrum shaping.

To retain the beam qualities from the laser accelerator and to be able to implement widespread applications, we have designed a Double Achromatic Beam Translation System (DABTS) based on weak-focusing magnets. We can achieve achromatic transmission in both the $x$ and $y$ directions, and also compress the bunch length. We reduce the influence of chromatic aberrations and high-order nonlinear terms with the design of fringe angles and special steering magnets.

## II. DESIGN OF DABTS

In a weak-focusing system with the field index $n$ between 0 and 1, the deflection angle $\theta$ will be larger than $2\pi$ if the transmissions for positon and divergence angle are both achromatic. After study, it is found that two specially designed weak-focusing magnets (M1 and M2) can make up a DABTS.

M1 and M2 both with the bending radius $r_c$ are the same except for deflecting direction. There is a drift space D1 (with length of $D_1$) between the beam source and M1; the drift space between M1 and M2 is D2 (with length of $D_2$); the length $D_3$ of drift space D3 between M2 and the exit is equal to $D_1$. Achromatic transmission in the $x$ and $y$ directions require these prerequisites: ① protons with the same energy have the same divergence angle in the $x$ direction and ② protons are all collimated in the $y$ direction between M1 and M2; ③ proton beams realize position-achromatic transmission in the $x$ direction from the target to the center of D2. Parallel transmission of each energy between M1 and M2 in the $x$ and $y$ directions makes the total transmission symmetrical about the center of D2, after which point the transmission is an inverse process, as shown in Fig. 1(b) and Fig. 1(e), enabling the transverse distributions at the exit (the end of D3) to return to the states at the beam source in theory, hence the beam size and transverse emittance can be retained. Achromatic transmission in both the $x$ and $y$ directions can be realized with this point-to-parallel-to-point optics; meanwhile, the pulse length can be compressed.

Under the requirements of ①, ② and ③, once $r_c$ and $D_1$ are identified, $(n, \theta)$, $D_2$, $D_3$ can be determined (see APPENDIX A and B). Generally there are two groups of $\theta$ ($< 2\pi$) satisfying the conditions. Then there are DABTS I and DABTS II, corresponding to the crossing point 1 and crossing point 2 in Fig. 5 respectively (see APPENDIX B).

### A. Beam transport in DABTS I

DABTS I can be designed with many combinations of $r_c$ and $D_1$. When $r_c = 0.15$ m, $D_1 = 0.343$ m, bunch length of 20 MeV proton beam is smaller; other parameters can be determined as ($n = 0.5$, $\theta = 5.226$ rad), $D_2 = 1.493$ m, $D_3 = D_1$.

In simulation, the proton beam has a central energy of 20 MeV and energy spread of 5%. The initial transverse distribution is a water-bag distribution with 10000 particles and the energy spectrum is an uniform distribution. The radius of the initial beam spot is set to 5 μm and the divergence angle is ±10 mrad. The root mean square (rms) sizes in the $x$ and $y$ directions are given by $x_{rms} = \sqrt{\overline{(x-\bar{x})^2}}$ and $y_{rms} = \sqrt{\overline{(y-\bar{y})^2}}$ respectively. The initial rms beam sizes are 1.77 μm in the $x$ and $y$ directions.

The envelope evolution is calculated with transformation matrixes 1.5 and 1.7 (see APPENDIX A), considering the influence of high-order nonlinear terms (namely $h \neq 0$ in 1.4 in APPENDIX A). The red, green, blue solid (dashed) curves represent envelopes with energy difference of 2.5%, 0%, -2.5% respectively in the $x$ ($y$) direction, as shown in Fig. 1(a). The three curves in the same color represent the envelopes of particles with initial divergence angle of 10, 0, -10 mrad respectively.

The beamline layout and the envelopes in the $x'z'$ plane in the laboratory frame of reference $x'y'z'$ is shown in Fig. 1(b). As laser accelerated beams are pulses, the envelopes can be crossed.

The proton distribution at the exit is shown in Fig. 1(c), and $x_{rms} = 28.9$ μm, $y_{rms} = 37.8$ μm. Most particles are inside a circular area of 100 μm in radius.

If the central energy changes and energy spread remains unchanged, the envelope and beam distribution will be the same without considering the influence of the space charge force. As the bending radius is pinned to 0.15 m, the magnetic field strength should vary corresponding to the central energy. For example, 100 MeV proton beams need 10 T, which can be achieved using high-field superconducting magnets, realizing significant reduction of dimensions and mass of magnets. Magnetic field strength can be lowered by increasing bending radius.

Energy spread and divergence angle lead to the transmission differences and cause an energy chirp in the beam longitudinal phase space. In DABTS, particles with higher energies have larger bending radii and longer trajectories, and particles at the bunch tail with lower energies have shorter trajectories. Hence longitudinal bunch can be compressed with proper designs. In this beamline, bunch length of 20 MeV proton beam is about 55 ps. For the convenience of discussion, the initial bunch length is set as 0 ps during the calculaliton. If the influence of high-order nonlinear terms is neglected, namely $h = 0$ in 1.4 in APPENDIX A, the bunch length will be 17 ps.

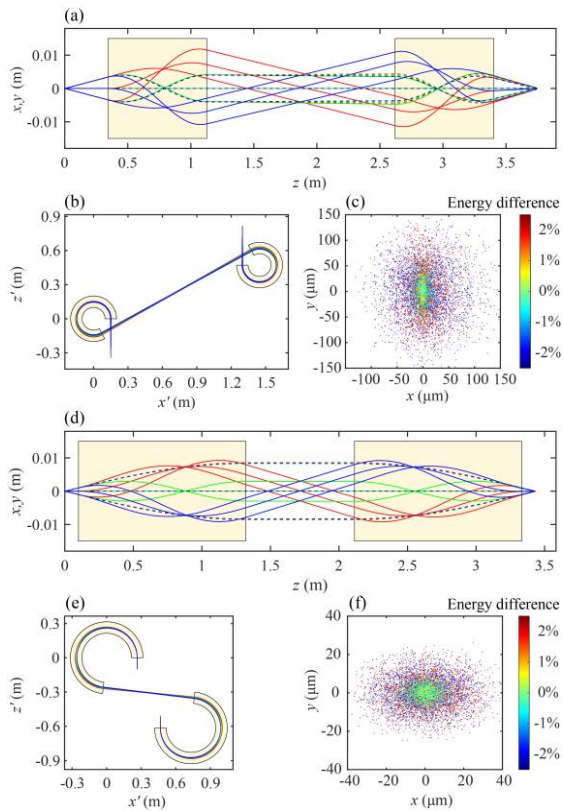

FIG. 1. DABTS I with $r_c = 0.15$ m and $D_1 = 0.343$ m: (a) The simulated envelope evolution of the proton beam with a central energy of 20 MeV and energy spread of 5%. The regions in a cornsilk background color denote M1 and M2. (b) The layout of the DABTS I in the $x'z'$ plane. Beam source is at $(r_c, -D_1)$. (c) Transverse distribution of the proton beam at the exit. DABTS II with $r_c = 0.265$ m and $D_1 = 0.1$ m: (d) The simulated envelope evolution of the proton beam with a central energy of 20 MeV and energy spread of 5%. (e) The layout of the DABTS II in the $x'z'$ plane. (f) Transverse distribution of the proton beam at the exit.

## B. Beam transport in DABTS II

The transmission in DABTS II has some differences with that in DABTS I. For example, $r_c = 0.265$ m, $D_1 = 0.1$ m, then we can get $(n = 0.1, \theta = 4.604)$. The envelope evolution is shown in Fig. 1(d). Differing from that in Fig. 1(a), envelope in the $y$ direction does not have focus points (beam waists) inside M1 and M2.

The beamline layout is shown in Fig. 1(e). In this type, the target chamber needs special design, as the beams in drift space D2 may pass through it. The laser is proposed to be incident on the target in the $x'y'$ plane.

Deflection angle is smaller in DABTS II than in DABTS I. Besides a cost savings, beam size is much smaller, as shown in Fig. 1(f), in which $x_{rms} = 10.6\ \mu\text{m}$, $y_{rms} = 6.4\ \mu\text{m}$. Most particles are inside a circular area of 15 μm in radius. Bunch length is 15.3 ps. If $h = 0$, the bunch length will be 7 ps.

Benefiting from the abilities of focusing beams in the $x$ and $y$ directions, and analyzing energy in the $x$ direction simultaneously, beams with large divergence angle and energy spread can be analyzed accurately in DABTS. After the transmission of certain deflection angle in M1, particles with different energies have been separated, and particles with the same energy and different initial angular divergence converge to the same focal point in the $x$ direction, where a slit can be placed to select energy, as shown in Fig. 1(a) and Fig. 1(d) in DABTS I and II respectively.

Beams in the $y$ direction are also at focal points at the slit in DABTS I, as the dashed curves show in Fig. 1(a). While in DABTS II, profiles are large and almost identical for different energies in the $y$ direction (Fig. 1(d)), which has more benefits for energy spectrum shaping, as particle number of each energy is selected in the *y* direction.

### C. Spatial and temporal properties of beams in DABTS

Spatial and temporal properties are important advantages of laser-accelerated beams, and many applications, such as WDM, time-resolved imaging and fast ignition, make great demands on them.

To investigate transmission characteristics systematically in DABTS, we scan parameters to find out the law of variations of spatial and temporal properties when $r_c$ and $D_1$ change. As mentioned above, once $r_c$ and $D_1$ are given, the whole beamline is determined in DABTS I or DABTS II. In the calculations, $r_c$ varies from 0.15 m to 2.05 m at an interval of 0.05 m; $D_1$ varies from 0.1 m to 1 m at an interval of 0.02 m. Some deflection angles $\theta$ may be close to $2\pi$ and not proper for applications, while these angles are also involved to comprehensively demonstrate the laws. The proton beam has a central energy of 20 MeV, energy spread of 5% and an initial divergence angle of ±10 mrad.

Figures 2(a) and 2(b) show the rms sizes in the *x* and *y* directions respectively at the exit in DABTS I. It is found that $r_c$ have a greater influence on sizes in the *x* direction compared with $D_1$. While the sizes in the *y* direction are largely decided by $D_1$. Small $r_c$ and $D_1$ are beneficial for obtaining small beam size. Figure 2(c) shows bunch lengths at the exit in DABTS I. Certain combination of $r_c$ and $D_1$ can minimize the bunch length of certain energy. For example, $r_c = 0.15$ m and $D_1 = 0.343$ m are optimal for 20 MeV proton beam.

Figures 2(d), 2(e) and 2(f) show rms sizes in the *x* and *y* directions and bunch length at the exit respectively in DABTS II. Beam sizes and bunch lengths are smaller in DABTS II than in DABTS I. The combination of $r_c = 0.265$ m and $D_1 = 0.1$ m minimizes the bunch length of 20 MeV proton beam.

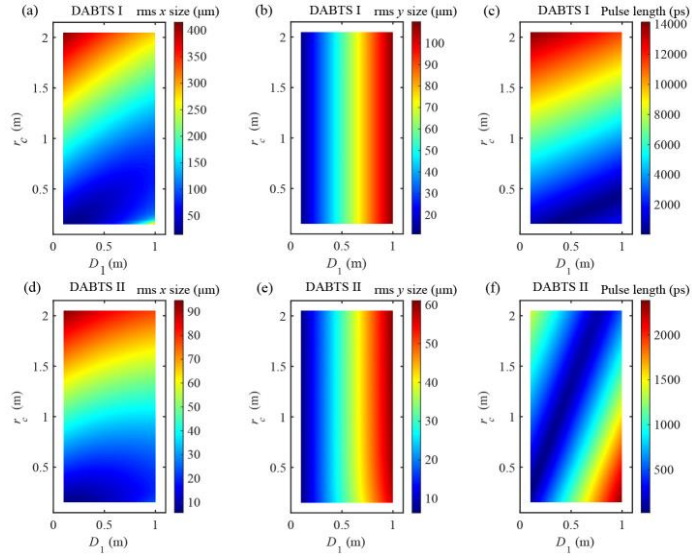

FIG. 2. Beam sizes and bunch lengths with different $r_c$ and $D_1$. (a) Beam rms size in the $x$ direction in DABTS I. (b) Beam rms size in the $y$ direction in DABTS I. (c) Bunch lengths in DABTS I. (d) Beam rms size in the $x$ direction in DABTS II. (e) Beam rms size in the $y$ direction in DABTS II. (f) Bunch lengths in DABTS II.

## III. INFLUENCE OF CHROMATIC ABERRATIONS AND IMPROVEMENT MEASURES

($n$, $\theta$) above are determined with 1.9 and 1.11 at central energy. Actually, deflection angle $\theta$ increases with beam energy and $r_c$ to fulfill ①, ② and ③. If $\theta$ are the same for different energies, influence of chromatic aberrations still exist, and deflection angle of low-energy particles (with negative energy difference) is larger than the angle which satisfies the requirements of ① and ②, leading to excessive focusing in the $x$ and $y$ directions and causing the longitudinal position of the image point to be smaller than that of central energy. For the similar reason, position of the image point of the high-energy particles (with positive energy difference) is larger due to less focusing. Then the whole beam cannot be transported with strict point-to-point optics, and the beam size at the image point of the central energy will increase with energy spread, but this is still far better than the case of transmission without achromatic design. Besides, the high-order nonlinear terms in 1.4 also have influences.

As protons have been roughly separated in the *x* direction at the exit of M1 and the entrance of M2, fringe angles of magnets can be designed to mitigate the impacts of chromatic aberrations and high-order nonlinear terms.

In the particle rest frame *zxy* whose origin (point *O*) is placed at the exit of M1, fringe angle $\alpha_1$ ($\alpha_2$) is the intersection angle between line *OA* (*BO*) and the positive *x*-axis in the *zx* plane, as shown in Fig. 3(a).

Protons with larger energies have larger radii and larger $\theta$ to realize achromatic transmission. Hence the slopes of lines *OA* and *BO* are supposed to be positive in Fig. 3(a). The slope *s* of line *OA* equals to $\tan\left(\frac{\pi}{2} - \alpha_1\right)$ (for line *BO*, $s = \tan\left(\frac{\pi}{2} - \alpha_2\right)$).

For an arbitrary particle, actual deflection angle $\theta$ is determined by its position $x_a$ in the *x* direction at the exit of M1; $\theta = \theta_0 + \varphi$; $\theta_0$ is the deflection angle of reference trajectory; $\varphi$ varies with $x_a$ and $\varphi = \arctan\left[\frac{x_a/s}{(r_c + x_a)}\right]$.

To realize an object-to-image point transport, particles with the same energy and different initial divergence angles should have the same deflection angle. After the adding of fringe angles, these particles will have different deflection angles, as they have different positions in the *x* direction at the exit of M1.

To keep the same total deflection angle for particles with the same energy and different initial divergence angles, special steering magnets (correction dipole magnet) is designed, with which divergence angle in the *x* direction will have a variation of $-\varphi$ as soon as particle leaves M1.

The shape of steering magnet in the *zx* plane is close to but not exact a combination of two sector magnets, as the regions in yellow show in Fig. 3(a). Length of steering magnet with a uniform magnetic field along $z$ axis is position-dependent in the $x$ direction, and proportional to $\varphi = \arctan\left[\dfrac{x_a/s}{(r_c + x_a)}\right]$. Deflecting directions in the two parts ($x_a > 0$ and $x_a < 0$) are supposed to be opposite if $\alpha_1$ and $\alpha_2$ are both positive. The part below ($x_a < 0$) is to compensate the reduction of deflection angle, and the part above ($x_a > 0$) is to resist the increase of deflection angle due to the fringe angle. Finally, total deflection angles are supposed to be the same for particles with the same energy in the $x$ direction to keep point-to-point imaging, and total deflection angles are not the same for particles with different energies to satisfy the requirements of ① and ②.

In practice, M1 and steering magnets can be designed as an integrated whole.

Symmetrical fringe angle and steering magnet are designed at the entrance of M2. The thin lens approximation is used to calculate the transmission in steering magnet.

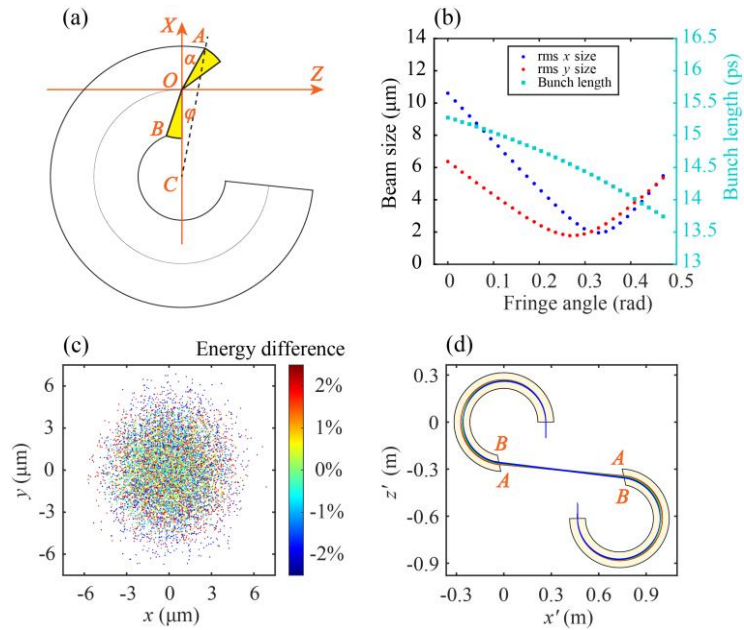

FIG. 3. Design of fringe angles and steering magnets and beam transmission in DABTS II with $r_c = 0.265$ m and $D_1 = 0.1$ m. (a) Design schematic of fringe angles and steering magnets. (b) The rms beam sizes in the $x$ and $y$

directions and bunch lengths at the exit varying with fringe angles in DABTS II. (c) Transverse distribution of proton beam with a central energy of 20 MeV, energy spread of 5% and an acceptance angle of ±10 mrad at the exit. (d) The layout of the DABTS II in the $x'z'$ plane. The deflection angle is 4.604 rad. The field index $n$ is 0.1.

Studies found that design of fringe angles and steering magnets can very effectively reduce the influence of chromatic aberrations. If the influence of high-order nonlinear terms is neglected, the beam transports are close to ideal achromatic transmission in both DABTS I and II with proper fringe angle $\alpha\,(=\alpha_1=\alpha_2)$.

If $h \neq 0$, the high-order nonlinear terms have different impacts in DABTS I and II. Figure 3(b) shows the rms beam sizes in the $x$ and $y$ directions and bunch lengths at the exit varying with fringe angles in DABTS II when $h \neq 0$. Fringe angles $\alpha_1 = \alpha_2 = 0.327$ rad can minimize the beam size and enable it to be kept on a micron scale, as shown in Fig. 3(c). Influence of chromatic aberrations and high-order nonlinear terms can be mitigated effectively in both the $x$ and $y$ directions. The rms beam sizes are 1.97 μm and 2.2 μm respectively in the $x$ and $y$ directions. Maximum beam sizes is less than 7 μm in both directions. The corresponding beamline layout is shown in Fig. 3(d), where the fringe angles at the exit of M1 and entrance of M2 are illustrated. Bunch length is 14.6 ps, which becomes shorter after the fringes angles are added.

For the same $r_c$ and $D_1$, the high-order nonlinear terms have larger impacts in DABTS I than in DABTS II, as $n$ is larger and focusing force is smaller in the $x$ direction in DABTS I, leading to larger envelopes and transmission differences in low-energy particles and high-energy particles. In this case, fringe angle $\alpha_1$ may need to be different from $\alpha_2$. For example with $\alpha_1 = 0.157$ rad and $\alpha_2 = -0.189$ rad, rms beam size in the $x$ direction can be shrunk from 29 μm in Fig. 1(c) to 9 μm under the influence of the high-order nonlinear terms, while the rms size in the *y* direction is enlarged from 38 μm to 43 μm; bunch length is 56 ps.

## IV. INFLUENCES OF ENERGY SPREAD AND DIVERGENCE ANGLE

DABTS with fringe angles design can effectively reduce the influence of chromatic aberrations, while the high-order nonlinear terms still have influence when energy spread and divergence angle increase. Under the influence of the high-order nonlinear terms, rms sizes in the *x* direction and bunch lengths increase dramatically with energy spread (2.5% ~ 30%) and divergence angle (±5 mrad ~ ±60 mrad) in DABTS I, as shown in Figs. 4(a) and 4(b) respectively. Fringe angles are set as $\alpha_1 = 0.157$ rad and $\alpha_2 = -0.189$ rad to reduce beam sizes.

The influences of the high-order nonlinear terms are much smaller in DABTS II when energy spread or divergence angle increases, as shown in Figs. 4(c) and 4(d) respectively. Fringe angles are set as $\alpha_1 = \alpha_2 = 0.327$.

As the high-order nonlinear terms do not impact transmission directly in the *y* direction, beam sizes are much smaller when energy spread and divergence angle increase.

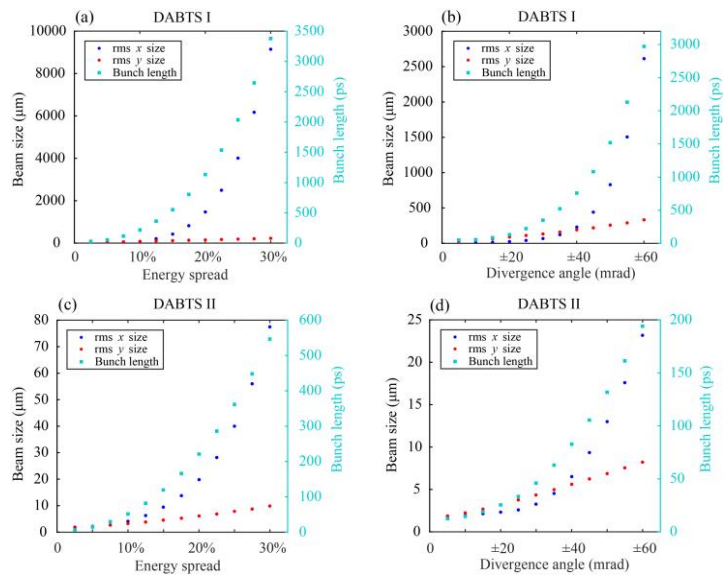

FIG. 4. Beam sizes and bunch lengths under different transmission conditions.

The central energy is 20 MeV. In DABTS I, $r_c = 0.15$ m, $D_1 = 0.343$ m: (a) Beam size and bunch length growth with energy spread when the acceptance angle is ±10 mrad. (b) Beam size and bunch length growth with divergence angle when the energy spread is 5%. In DABTS II, $r_c = 0.265$ m, $D_1 = 0.1$ m: (c) Beam size and bunch length growth with energy spread when the acceptance angle is ±10 mrad. (d) Beam size and bunch length growth with divergence angle when the energy spread is 5%.

A large area of uniformly distributed particles is required in many irradiation applications, such as cancer therapy. Combination of DABTS and quadrupole triplet lens can realize energy spectrum shaping and point-to-parallel-to-point-to-parallel transport simultaneously for irradiation applications.

## V. CONCLUSION

In summary, to apply ion beams with a micron scale and picosecond time scale in a laser accelerator, two types of DABTS (I and II) were designed using point-to-parallel-to-point optics, and achromatic transmission in both the horizontal and vertical directions and compression of bunch length were achieved simultaneously based on weak-focusing magnets. The influences of parameters in DABTS were investigated to find out the minimum beam size and bunch length. To realize close to ideal point-to-parallel-to-point optics, fringe angles and special steering magnets were designed to effectively reduce the influence of chromatic aberrations and high-order nonlinear terms. The influences of energy spread and divergence angle were discussed, and it was found that the high-order nonlinear terms have much larger impact in DABTS I than in DABTS II. Beams with large divergence angle and energy spread can be analyzed accurately in DABTS, then energy spectrum shaping can be achieved. We believe that our works have systematically solved the transport problems of beams with large energy spread and divergence angle in laser accelerators. With DABTS, laser accelerators will be competent to a wide range of applications, such as WDM, fast ignition, high time-resolved imaging of fast transient phenomena, etc.

## ACKNOWLEDGMENTS

This work has been supported by Fundamental Research Program of Shenzhen (SZWD2021007, JCYJ20200109105606426), National Natural Science Foundation of China (92050203), Science and Technology on Plasma Physics Laboratory. The authors are very grateful to Matthew J Easton for spending much time and energy making careful revisions to our paper, and the help from Dongyu Li, Kun Zhu, Xueqing Yan and Chen Lin at Peking University.